\pdfoutput=1
\newcommand*{\ATLASLATEXPATH}{latex/}

\documentclass[UKenglish,texlive=2016,paper=letter,cernpreprint]{\ATLASLATEXPATH atlasdoc}

\usepackage[backend=biber]{\ATLASLATEXPATH atlaspackage}
\usepackage{\ATLASLATEXPATH atlasbiblatex}
\usepackage{\ATLASLATEXPATH atlascontribute}
\usepackage{\ATLASLATEXPATH atlasphysics}
\graphicspath{{logos/}}
\usepackage{rotating}
\usepackage{overpic}
\usepackage{multirow}
\usepackage{tabb}

\newcommand\dbline{\noalign{\vskip 0.10truecm\hrule\vskip 0.05truecm\hrule\vskip 0.10truecm}}
\newcommand\sgline{\noalign{\vskip 0.10truecm\hrule\vskip 0.10truecm}}
\definecolor{myg}{RGB}{178,164,108}
\definecolor{myb}{RGB}{28,41,87}

\newcommand{\asp}[1]{$\!\!$#1$\!\!$}
\newcommand{\bsp}{\!\!\!}
\newcommand{\Bsp}{\!\!\!\!}

\newcommand{\BSP}{\!\!\!\!\!\!\!\!\!\!\!\!}
\newcommand{\HTmiss}{H_\text{T}^\text{miss}}
\newcommand{\ETmiss}{E_\text{T}^\text{miss}}
\newcommand{\ETmissvec}{\vec{E}_\text{T}^\text{miss}}
\newcommand{\smallZ}{{\mbox{\tiny\textit{Z}}}}
\newcommand{\smallW}{{\mbox{\tiny\textit{W}}}}

\renewcommand{\pT}{p_\text{T}}
\renewcommand{\muR}{\mu_{\text{R}}}
\renewcommand{\muF}{\mu_{\text{F}}}
\renewcommand{\MeV}{\,\text{Me\kern -0.1em V}}
\renewcommand{\GeV}{\,\text{Ge\kern -0.1em V}}
\renewcommand{\TeV}{\,\text{Te\kern -0.1em V}}
\renewcommand{\ifb}{\,\mbox{fb$^{-1}$}}
\renewcommand{\ipb}{\,\mbox{pb$^{-1}$}}
\newcommand{\pb}{\,\mbox{pb}}
\definecolor{mygold}{RGB}{178,164,108}
\definecolor{myblue}{RGB}{28,41,87}

\hypersetup{pdftitle={ATLAS document},pdfauthor={The ATLAS Collaboration}}

\PreprintIdNumber{CERN-EP-2018-184}
\AtlasTitle{Search for invisible Higgs boson decays in vector boson fusion at $\sqrt{s} = 13\TeV$ with the ATLAS detector}
\AtlasAbstract{
  We report a search for Higgs bosons that are produced via vector boson fusion and subsequently decay into invisible particles.
  The experimental signature is an energetic jet pair with invariant mass of $\mathcal{O}(1)\TeV$ and $\mathcal{O}(100)\GeV$ missing transverse momentum.
  The analysis uses $36.1\ifb$ of $pp$ collision data at $\sqrt{s}{\,=\,}13\TeV$ recorded by the ATLAS detector at the LHC.
  In the signal region the $2252$ observed events are consistent with the background estimation.
  Assuming a $125\GeV$ scalar particle with Standard Model cross sections,
  the upper limit on the branching fraction of the Higgs boson decay into invisible particles is $0.37$ at $95\%$ confidence level where $0.28$ was expected.
  This limit is interpreted in Higgs portal models to set bounds on the \textsc{wimp}--nucleon scattering cross section.
  We also consider invisible decays of additional scalar bosons with masses up to $3\TeV$
  for which the upper limits on the cross section times branching fraction are in the range of $0.3$--$1.7\pb$.
}
\AtlasRefCode{EXOT-2016-37}
\AtlasJournalRef{Phys.\ Lett.\ B 793 (2019) 499}

\begin{document}
\maketitle

\section{Introduction}

We present a search for the decays of the Higgs boson \cite{ATLAS:2012af,Chatrchyan:2012ia}, produced via the vector boson fusion (VBF) process \cite{LHCHiggsCrossSectionWorkingGroup:2011ti,LHCHiggsCrossSectionWorkingGroup:2012vm},
into invisible particles ($\chi\bar\chi$) with an anomalous and sizable $\mathcal{O}(10)\%$ branching fraction.
The hypothesis under consideration 
\cite{Antoniadis:2004se,ArkaniHamed:1998vp,Kanemura:2010sh,Djouadi:2011aa,Shrock:1982kd,Choudhury:1993hv,Eboli:2000ze,Davoudiasl:2004aj,Godbole:2003it,Ghosh:2012ep,Belanger:2013kya,Curtin:2013fra}
is that the Higgs boson might decay into a pair of weakly interacting massive particles (\textsc{wimp}) \cite{Goldberg:1983nd,Ellis:1983ew},
which may explain the nature of dark matter (see Ref.~\cite{Clowe} and the references therein).
The search carried out for the $125\GeV$ Higgs boson is repeated for hypothetical scalars with masses up to $3\TeV$.
The search is independent on the decay of the mediator because the final state particles are invisible to the detector,
while it is dependent on its $\ETmiss$ distribution (defined below) because that quantity is reflective of the mediator's $\pT$ distribution.

The data sample corresponds to an integrated luminosity of $36.1\ifb$ of proton-proton ($pp$) collisions at $\sqrt{s}{\,=\,}13\TeV$ recorded by the ATLAS detector at the LHC in 2015 and 2016.
The experimental signature of the VBF production process is a pair of energetic quark jets with a wide gap in pseudorapidity ($\eta$)
corresponding to the $\mathcal{O}(1)\TeV$ value of the invariant mass ($m_{jj}$) of the highest-$\pT$ jets in the event.\footnote{%
  ATLAS uses a right-handed coordinate system with its origin at the nominal interaction point in the center of the detector and the $z$-axis along the beam direction.
  The $x$-axis points from the interaction point to the center of the LHC ring;
  the $y$-axis points upward.
  Cylindrical coordinates $(r,\phi)$ are used in the transverse plane, where $\phi$ is the azimuthal angle around the $z$-axis.
  The pseudorapidity is defined as $\eta{\,=\,}-\ln(\tan{(\theta/2)})$, where $\theta$ is the polar angle.
}
The signature for the decay process is the $\mathcal{O}(100)\GeV$ value of the missing transverse momentum $\big(\ETmiss\big)$ that corresponds to the Higgs boson $\pT$.
The VBF topology offers a powerful rejection of the strongly produced\footnote{%
  For the $W$ and $Z$ background processes in this paper, electroweak (EW) refers to diagrams that are of $\mathcal{O}(\alpha_{\textsc{ew}}^4)$ or greater,
  while strong refers to diagrams that are of $\mathcal{O}(\alpha_{\textsc{s}}^2)$ or greater accompanied by $\mathcal{O}(\alpha_{\textsc{ew}}^2)$.
}
backgrounds due to single vector boson plus two jets, and the multijet background produced from QCD processes.
In this analysis, the Higgs production via the gluon fusion mechanism is subdominant to VBF and is considered as part of the signal.

Direct searches for invisible Higgs decays look for an excess of events over Standard Model expectations.
The absence of an excess is interpreted as an upper limit on the branching fraction of invisible decays ($\mathcal{B}_\text{inv}$) assuming the Standard Model production cross section \cite{deFlorian:2016spz} of the $125\GeV$ Higgs boson.
Other published results have targeted a variety of production mechanisms---%
gluon fusion, VBF, $W$ or $Z$ associated production
\cite{Aad:2014wza,Chatrchyan:2014tja,atlas:VjjHinv,Khachatryan:2016mdm,Aaboud:2017bja}%
---to set upper limits on $\mathcal{B}_\text{inv}$.
The best limits are from the statistical combination of search results for which
ATLAS reports an observed (expected) limit of $0.26$ ($0.17$) \cite{ATLAS:2019inv} and
CMS reports $0.26$ ($0.20$) \cite{Sirunyan:2018owy} at $95\%$ confidence level (CL).
For these combinations the single input with the highest expected sensitivity is VBF,
the channel pursued here.
For the VBF channel using Run-1 data, ATLAS reports $0.28$ ($0.31$) \cite{Aad:2015txa} and CMS reports $0.43$ ($0.31$) \cite{Khachatryan:2016whc}.
In a more recent update of the VBF channel using Run-2 data, ATLAS reports $0.37$ ($0.28$) [this paper] CMS reports $0.33$ ($0.25$) \cite{Sirunyan:2018owy}.

Global fits to the measurements of visible decay channels of the Higgs boson place indirect constraints on the beyond-the-SM decay branching fraction $\mathcal{B}_{\textsc{bsm}}$.
The $\mathcal{B}_{\textsc{bsm}}$ is the \textit{sum} of
$\mathcal{B}_\text{inv}$ that represents invisible decays and
$\mathcal{B}_\text{undet}$ that represents the channels that are undetected,
i.e.,
those that are not included in the following combination.
For $\mathcal{B}_\textsc{bsm}$ using Run-1 data,
ATLAS reports $0.49$ ($0.48$) \cite{Aad:2015pla} and
CMS reports $0.57$ ($0.52$) \cite{Khachatryan:2014jba} with similar but not identical assumptions.
A combination of ATLAS and CMS results using Run-1 data gives $0.34$ ($0.39$) \cite{Khachatryan:2016vau}. 
In a more recent update using Run-2 data, CMS reports an observed limit on $\mathcal{B}_\text{undet}$ of $0.38$ \cite{Sirunyan:2018koj}.
As noted in Ref.~\cite{Aad:2015txa},
there is complementarity between the direct search for invisible Higgs decays and the indirect constraints from the global fits.

In this analysis, several changes and improvements are made with respect to the previous ATLAS paper on this topic \cite{Aad:2015txa}.
The trigger and hadronic objects are defined considering the simultaneous $pp$ collisions in the same and nearby bunch crossings (pileup) (Section~\ref{sec:obj}).
The leading backgrounds are simulated using state-of-the-art QCD predictions (Section~\ref{sec:sim}).
The event selections are changed to retain a good sensitivity despite the higher pileup.
The analysis extracts the signal yield using a binned likelihood fit to the $m_{jj}$ spectrum in $3$ bins to increase the signal sensitivity (Section~\ref{sec:sr}). 
The estimation of the important and dominant background for the $Z{\,\rightarrow\,}{\nu\nu}$ process (denoted $Z_{\nu\nu}$) relies only on the $Z_{ee}$ and $Z_{\mu\mu}$ control samples,
and is not affected by theoretical uncertainties of the $W$-to-$Z$ extrapolation (Section~\ref{sec:cr}).
The systematic uncertainties are evaluated separately for each $m_{jj}$ bin (Section~\ref{sec:syst}).
The search is repeated for other scalars with masses up to $3\TeV$,
which can easily be reinterpreted for models not considered in this Letter (Section~\ref{sec:results}).
Several aspects of the analysis have not changed compared to the ATLAS Run-1 analysis---e.g.,
subdetector descriptions, transfer factor method, Higgs portal models---%
and details of these may be found in Ref.~\cite{Aad:2015txa}.

\section{Detector, trigger, and analysis objects}\label{sec:obj}

ATLAS is a multipurpose particle physics detector with a forward--backward symmetric cylindrical geometry consisting of a tracking system, electromagnetic and hadronic calorimeters, and a muon system \cite{Aad:2008zzm}.

The trigger to record events in the sample containing the VBF signal candidates used a two-level $\ETmiss$ algorithm with thresholds adjusted throughout the data-taking period to cope with varying levels of pileup \cite{ATLAS-CONF-2014-002,Hamano:2275659}.
The \textsc{level-}{\footnotesize 1} system used coarse-granularity analog sums of the energy deposits in the calorimeter towers to require $\ETmiss{\,>\,}50\GeV$.
The second-level \textsc{high level trigger} system \cite{ATLASTDAQ:2016pov} used jets that are reconstructed from calibrated clusters of cell energies \cite{Aad:2016upy} and requires $\ETmiss{\,>\,}70$--$110\GeV$ depending on the luminosity and the pileup level.
The trigger efficiency \cite{Aaboud:2016leb} for signal events is $98\%$ for $\ETmiss{\,>\,}180\GeV$ when comparing the trigger selection with the offline $\ETmiss$ definition that contains additional corrections.

The triggers to record the control samples for background studies used lepton and jet algorithms \cite{ATL-DAQ-PUB-2017-001}.
The samples with leptonic $W$ and $Z$ decays were collected with a single-electron or -muon trigger with $\pT{\,>\,}24\GeV$ ($26\GeV$) and an isolation requirement in 2015 (towards the end of 2016).
The sample of multijet events was collected using a set of low-threshold single-jet triggers with large prescale values to keep the event rate relatively low.

For each event, a vertex is reconstructed from two or more associated tracks ($t$) with $\pT{\,>\,}400\MeV$.
If multiple vertices are present, we consider the one with the largest $\sum_t(p_{\text{T},t})^2$ as the primary vertex of our candidates.

Leptons ($\ell{\,=\,}e,\mu$) are identified to help characterize events with leptonic final states from decays of vector bosons.
Since the signal process contains no leptons, such events are used for the background estimation, which is described in Section~\ref{sec:cr}.
Electrons (muons) must have $\pT{\,>\,}7\GeV$, $|\eta|{\,<\,}2.47$ ($2.5$), and satisfy an isolation requirement.
Electrons are reconstructed by matching clustered energy deposits in the electromagnetic calorimeter to tracks from the inner detector \cite{Aaboud:2016vfy,ATLAS-CONF-2016-024}
and muons by matching inner detector and muon spectrometer tracks \cite{Aad:2016jkr}.
For electrons (muons) with a $\pT$ value of at least $30\GeV$ ($20$-$100\GeV$), the reconstruction efficiency $80\%$ ($96\%$) with a rejection factor of around $500$ ($600$).
All leptons must originate from the primary vertex.

Jets are reconstructed from topological clusters in the calorimeters using the anti-$k_t$ algorithm \cite{AntiKt} with a radius parameter $R{\,=\,}0.4$.
Jets must have $\pT{\,>\,}20\GeV$ and $|\eta|{\,<\,}4.5$.
The subset of jets with $\pT{\,<\,}60\GeV$ and $|\eta|{\,<\,}2.4$ are jet vertex tagged (\textsc{jvt}) \cite{Aad:2015ina} to suppress pileup effects, using tracking and vertexing.
The \textsc{jvt} is $92\%$ efficient for the jets in the signal process from the primary interaction with a rejection factor of around $100$ for pileup jets with $\pT$ value in the range of $20$-$50\GeV$ \cite{Aad:2015ina}.

Cleaning requirements help suppress non-collision backgrounds \cite{ATLAS-CONF-2015-029}.
Fake jets due to noisy cells are removed by requiring a good fit to the expected pulse shape for each constituent calorimeter cell.
Fake jets induced by beam-halo interactions with the LHC collimators are removed by requirements on their energy distribution and the fraction of their constituent tracks that originate from the primary vertex.

In events with identified leptons, an overlap removal procedure is applied to resolve the ambiguities in cases where a jet is also identified in the same $\eta$-$\phi$ area,
which could occur in situations such as having a heavy-flavor hadron decay within a jet \cite{Aaboud:2018mna}.
The lepton--jet overlap in $\Delta{R}$ distance\footnote{%
The distance variable is defined as $\Delta{R}{\,=\,}\sqrt{(\Delta{\eta})^2{\,+\,}(\Delta{\phi})^2}$.}
is resolved sequentially as follows.
If an electron is near a jet with $\Delta{R}{\,<\,}0.2$, the jet is removed to avoid the double counting of electron energy deposits.
If a remaining jet is near an electron with $0.2{\,\le\,}\Delta{R}{\,<\,}0.4$, the electron is removed.
If a muon is near a jet with $\Delta{R}{\,<\,}0.4$ and the jet is associated with at least (less than) three charged tracks with $\pT{\,>\,}500\MeV$, the muon (jet) is removed.

The $\ETmiss$ variable is the magnitude of the negative vector sum of the transverse momenta, $-\sum_i\vec{p}_{\text{T},i}$,
where $i$ represents both the ``hard objects'' and the ``soft term.''
The hard objects consist of leptons and jets, which are individually reconstructed and calibrated;
the list excludes pileup jets, 
which are removed by a \textsc{jvt} requirement.
The soft term is formed from inner detector tracks not associated with the hard objects, but matched to the primary vertex.
In the search region, the $\ETmiss$ produced by the Higgs decay is balanced in the transverse plane by the dijet system.

The \textsc{jvt} procedure is intended to remove pileup jets, but can cause large fake $\ETmiss$ if it removes a high-$\pT$ jet from the hard scatter, 
e.g., a jet from a $\pT$-balanced three-jet event.
In order to reduce this, a correlated quantity $\HTmiss$---defined as $|\sum_j\vec{p}_{\text{T},j}|$,
where $j$ represents all jets without the \textsc{jvt} requirement---is required to be $\HTmiss{\,>\,}150\GeV$. 
In the three-jet example, $\HTmiss$ would be near zero.

The $\ETmiss$ significance ($S_\textsc{met}$) is used only in events with one identified electron in the final state and is defined
as $\ETmiss/\sqrt{p_{\text{T},j_1}{+}p_{\text{T},j_2}{+}p_{\text{T},e}}$,
where the $\pT$ quantities are for leading jet ($j_1$), subleading jet ($j_2$), and electron, respectively.
The use of this quantity to reduce the contamination from jets misidentified as electrons is discussed in Section~\ref{sec:cr}.

\section{Event simulation}\label{sec:sim}

Monte Carlo simulation (MC) consists of an event generation followed by detector simulation \cite{simuAtlas} using \textsc{geant}{\footnotesize 4} \cite{GEANT4}.
Simulated events were corrected for the small differences between data and MC in the trigger, the lepton identification efficiency, and the jet energy scale and resolution using dedicated data samples.

For the signal process, the VBF events were generated at next-to-leading order (NLO) in QCD using \textsc{powheg-box}{\footnotesize 2} \cite{powheg5};
inclusive NLO electroweak corrections were applied using \textsc{hawk} \cite{hawk}.
The generated events were interfaced with \textsc{pythia}{\footnotesize 8} \cite{pythia81} for hadronization and showering,
using the \textsc{aznlo} tune \cite{Aad:2014xaa} and the \textsc{nnpdf}{\footnotesize 3.0} NNLO PDF set \cite{Ball:2014uwa}.
The gluon fusion events were generated using \textsc{powheg-nnlops} \cite{Hamilton:2013fea}
with the \textsc{pdf}{\footnotesize 4}\textsc{lhc}{\footnotesize 15} PDF set \cite{Butterworth:2015oua} 
interfaced to a fast detector simulation \cite{1742-6596-396-2-022031,ATLAS:2018ghb,ATLAS:2018sim}.
The cross section for ggF (VBF) was computed at N$^3$LO (NNLO) in QCD and NLO (NLO) in electroweak.
The showering simulation followed the same procedure as for the VBF sample.
For both the VBF and gluon fusion events, the $H{\,\rightarrow\,}ZZ^\ast{\,\rightarrow\,}4\nu$ process is included in the sample as invisible decays of the Higgs boson.
Additional scalars with masses up to $3\TeV$ were simulated as described above for VBF signal process, assuming a full width of $4\MeV$.

The $W$ and $Z$ events were generated using \textsc{sherpa}{\footnotesize 2.2.1} \cite{Gleisberg:2008ta}
with \textsc{comix} \cite{Gleisberg:2008fv} and \textsc{openloops} \cite{Cascioli:2011va} matrix-element generators,
and merged with \textsc{sherpa} parton shower \cite{Schumann:2007mg} using the \textsc{me}{\footnotesize +}\textsc{ps}{\footnotesize @}\textsc{nlo} prescription \cite{Hoeche:2012yf}.
The \textsc{nnpdf}{\footnotesize 3.0} NNLO PDF set was used.
In terms of the order of the various processes,
the strong production was calculated at NLO for up to two jets and leading order (LO) for the third and fourth jets.
The electroweak production was calculated at LO for the second and third jets.
The levels of the interference between electroweak and strong processes were computed with \textsc{madgraph}{\footnotesize 5}\_a\textsc{mc}{\footnotesize @}\textsc{nlo} \cite{Alwall:2014hca}.
The interference on the total expected background is only $0.1\%$ and thus neglected. 

Other potential background processes involve top quarks, dibosons, and multijets.
Top quarks and dibosons were generated with \textsc{powheg} interfaced with \textsc{pythia}
and \textsc{evtgen} \cite{Lange:2001uf}, which simulate the heavy-flavor decays.
The diboson backgrounds include electroweak-mediated processes.
The multijet estimate does not directly use the MC.

To each hard-scatter MC event, pileup collisions ($30$ on average) were added to mimic the environment of the LHC. 
The pileup collisions, simulated with \textsc{pythia}{\footnotesize 8} \cite{pythia81} using \textsc{mstw}{\footnotesize 2008} PDF \cite{Martin:2009iq} and the \textsc{a}{\footnotesize 2} set of tuned parameters \cite{AU2},
were subsequently reweighted to reproduce the pileup distribution in data.

The sizes of the MC samples vary depending on the process.
The effective luminosity ranges for the MC samples varies depending on the process and on the selections, which are defined in Section~\ref{sec:sr}.
For the $W$ process, the MC sample is approximately half of that of the data selected for the $W$ control region and also half for the signal region.
For the $Z$ process, the MC sample for the $Z_{\ell\ell}$ subprocess is approximately twice that of data in the $Z$ control region;
the MC sample for $Z_{\nu\nu}$ subprocess is approximately the same as that of data in the signal region.

\section{Event selection}\label{sec:sr}

All events must have a primary vertex.
The selection listed below divides the data sample into a signal-enriched search region (SR) and background-enriched control regions (CR).
The control regions and the statistical fit are discussed in detail in Section~\ref{sec:cr}.
The rest of this section focuses on the SR and the prefit event yields.\footnote{%
  ``Prefit'' indicates that the event yields are not adjusted according to the statistical treatment of the background predictions,
  which is described in the second half of Section~\ref{sec:cr}.
  ``Postfit'' label the the quantities that come out of the fit procedure.
}

For the SR, an event is required to have 
\begin{itemize}
  \setlength\itemsep{0pt}
  \item no isolated electron or muon,
  \item a leading jet with $\pT{\,>\,}80\GeV$,
  \item a subleading jet with $\pT{\,>\,}50\GeV$,
  \item no additional jets with $\pT{\,>\,}25\GeV$,
  \item $\ETmiss{\,>\,}180\GeV$,
  \item $\HTmiss{\,>\,}150\GeV$.
\end{itemize}
The two jets are required to have the following properties:
\begin{itemize}
  \setlength\itemsep{0pt}
  \item not be aligned with $\ETmissvec$, $|\,\Delta\phi_{j\text{-}\textsc{met}}\,|{\,>\,}1$,
  \item not be back-to-back, $|\,\Delta\phi_{jj}\,|{\,<\,}1.8$,
  \item be well separated in $\eta$, $|\,\Delta\eta_{jj}\,|{\,>\,}4.8$,
  \item be in opposite $\eta$ hemispheres, $\eta_{j_1}{\,\cdot\,}\eta_{j_2}{\,<\,}0$,
  \item $m_{jj}{\,>\,}1\TeV$.
\end{itemize}
The SR includes background events containing a $W$ or $Z$ plus two jets, where 
the $W$ decays into $e\nu$, $\mu\nu$, and $\tau\nu$, and
the $Z$ decays into two neutrinos.
Here the leptons from the $W$ decays are not reconstructed since they would otherwise be rejected by the selection.

Table~\ref{tab_01} gives the prefit SR yields in the first column.
The VBF production process gives the biggest contribution ($87\%$) to the signal sample (fixed as $\mathcal{B}_\text{inv}{\,=\,}1$).
The contribution from gluon fusion accompanied by parton radiation is small ($13\%$) and other production modes contribute negligibly.
The fraction of VBF signal events that pass the signal region event selections,
defined as acceptance times reconstruction efficiency,
is $0.7\%$.
As is discussed in Section~\ref{sec:results},
the signal significance is improved by considering three bins of $m_{jj}$ defined as follows:
$1{\,<\,}m_{jj}{\,\le\,}1.5\TeV$,
$1.5{\,<\,}m_{jj}{\,\le\,}2\TeV$, and
$m_{jj}{\,>\,}2\TeV$.
The prefit $S/B$ ratio (for $\mathcal{B}_\text{inv}{\,=\,}1$) in these bins is approximately $0.3$, $0.4$, $0.8$, respectively.

For the backgrounds, both the strong production and the EW production contribute in the SR.
The strong production processes contributes more than $70\%$ of the backgrounds in all of the $m_{jj}$ bins.
There is variation in the EW fractions for the background processes due to a combination of the following factors:
known differences in the production diagrams between $W$ and $Z$,
differences in kinematic acceptance for the particular $W$ or $Z$ decay,
and differences in the MC sample size for each EW process.

\begin{table}[bt!]
{\small
\caption{
  Event yields in the signal region (SR) and control regions (CR) summed over lepton charge and flavor.
  The yields are the \textit{prefit} values for $m_{jj}{\,>\,}1\TeV$.
  The observed data ($N$), the background estimate ($B$), and the signal ($S$ for $m_{H}{\,=\,}125\GeV$ with $\mathcal{B}_\text{inv}{\,=\,}1$) are given.
  The $B$ and $S$ values for individual processes are rounded to a precision commensurate with the sampling uncertainty associated with the finite MC sample size.
  For all processes the fractions of electroweak production [\textsc{ew}] are given.
  ``Other'' is defined in the text.
}%
\label{tab_01}
\begin{center}
\begin{tabular*}{0.55\columnwidth}{l rl p{0.00\columnwidth} rl p{0.00\columnwidth} rl}
  \dbline
  \multicolumn{1}{l}{Description\Bsp}
  &  \multicolumn{2}{c}{\Bsp SR}
  && \multicolumn{2}{c}{\Bsp $W$ CR}
  && \multicolumn{2}{c}{\Bsp $Z$ CR}
  \\
  \cline{2-3}\cline{5-6}\cline{8-9}%
  \multicolumn{1}{l}{$\large\phantom{X^X}$\Bsp}
  &  \multicolumn{1}{r}{\Bsp Yield} &\Bsp[\textsc{ew}]
  && \multicolumn{1}{r}{\Bsp Yield} &\Bsp[\textsc{ew}]  
  && \multicolumn{1}{r}{\Bsp Yield} &\Bsp[\textsc{ew}] \\
  \sgline
  \multicolumn{1}{l}{$N$, observed}                        &\Bsp 2252 &\Bsp                  \Bsp&&\Bsp 1602 &\Bsp                  \Bsp&&\Bsp 166 &\Bsp        \Bsp\\
  \multicolumn{1}{l}{$B$, expected}                        &\Bsp 2243 &\Bsp                  \Bsp&&\Bsp 1648 &\Bsp                  \Bsp&&\Bsp 183 &\Bsp        \Bsp\\
  \multicolumn{1}{l}{\quad $Z{\,\rightarrow\,}\nu\nu$}     &\Bsp 1111 &\Bsp [18\%]           \Bsp&&\Bsp -    &\Bsp                  \Bsp&&\Bsp -   &\Bsp        \Bsp\\
  \multicolumn{1}{l}{\quad $Z{\,\rightarrow\,}ee,\mu\mu$}  &\Bsp 12   &\Bsp [\phantom{1}9\%] \Bsp&&\Bsp 38   &\Bsp [\phantom{1}9\%] \Bsp&&\Bsp 181 &\Bsp [23\%] \Bsp\\
  \multicolumn{1}{l}{\quad $Z{\,\rightarrow\,}\tau\tau$}   &\Bsp 10   &\Bsp [16\%]           \Bsp&&\Bsp 11   &\Bsp [16\%]           \Bsp&&\Bsp -   &\Bsp        \Bsp\\
  \multicolumn{1}{l}{\quad $W{\,\rightarrow\,}e\nu,\mu\nu$}&\Bsp 540  &\Bsp [16\%]           \Bsp&&\Bsp 1400 &\Bsp [30\%]           \Bsp&&\Bsp -   &\Bsp        \Bsp\\
  \multicolumn{1}{l}{\quad $W{\,\rightarrow\,}\tau\nu$ }   &\Bsp 533  &\Bsp [20\%]           \Bsp&&\Bsp 130  &\Bsp [34\%]           \Bsp&&\Bsp -   &\Bsp        \Bsp\\
  \multicolumn{1}{l}{\quad Other}                          &\Bsp 36   &\Bsp                  \Bsp&&\Bsp 67   &\Bsp                  \Bsp&&\Bsp 2   &\Bsp        \Bsp\\
  \sgline
  \multicolumn{1}{l}{$S$, signal}                          &\Bsp 1070 &\Bsp                  \Bsp&&\Bsp -    &\Bsp                  \Bsp&&\Bsp -   &\Bsp        \Bsp\\
  \multicolumn{1}{l}{\quad VBF}                            &\Bsp 930  &\Bsp                  \Bsp&&\Bsp -    &\Bsp                  \Bsp&&\Bsp -   &\Bsp        \Bsp\\
  \multicolumn{1}{l}{\quad Gluon fusion}                   &\Bsp 140  &\Bsp                  \Bsp&&\Bsp -    &\Bsp                  \Bsp&&\Bsp -   &\Bsp        \Bsp\\
  \dbline
\end{tabular*}
\end{center}
}
\end{table}

\section{Control samples and statistical treatment}\label{sec:cr}

The main backgrounds in the SR, comprising of $98\%$ of the background, are the $W$ and $Z$ processes.
The minor backgrounds, comprising the remaining $2\%$, are the diboson, $t\bar{t}$, and multijet processes.
Accurate estimation of the $W$ and $Z$ processes is the biggest challenge of the analysis.
The main background yields are extracted using dedicated control samples in data.

This section is organized as follows.
First, the two main CR are described and the associated prefit yields are given.
Second, the fit parameters are defined along with a discussion of the contamination in the $W_{e\nu}$ subsample.
Third, the fit procedure is described and the postfit yields are stated.
Lastly, the minor backgrounds and the estimation of the multijet processes are described.

The $W$ CR requires one identified lepton with a $\pT$ threshold of $30\GeV$, but the selections are otherwise identical to those of the SR.
The initial $\ell\nu$ selection is divided by lepton flavor, charge, and, for the $e\nu$ final state,
a passing selection on $S_\textsc{met}{\,>\,}4\sqrt{\!\GeV}$ to define four $W$ CR subsamples
$\big(W_{\mu^+\nu}$, $W_{\mu^-\nu}$, $W_{e^+\nu}^\textsc{high}$, $W_{e^-\nu}^\textsc{high}\big)$.
The complementary failed selection on $S_\textsc{met}$ defines the two ``fake-enriched'' subsamples
$\big(W_{e^+\nu}^\textsc{low}$, $W_{e^-\nu}^\textsc{low}\big)$.
The $\ETmiss$ is calculated by adding the calibrated leptons to the sum.

The $Z$ CR is based on the same selection criteria as the SR, but the lepton veto is replaced by the requirement of two same-flavor opposite-sign leptons $\ell$ with $|m_{\ell\ell}-m_{Z}|<25\GeV$.
The sample is divided by lepton flavor, but not by charge ($Z_{ee}$, $Z_{\mu\mu}$).
The leading lepton-$\pT$ threshold is the same as above, and the subleading lepton-$\pT$ threshold is $7\GeV$. 
The $\ETmiss$ is calculated as is done above.

Table~\ref{tab_01} gives the prefit CR yields for the inclusive selection of $m_{jj}{\,>\,}1\TeV$ for the $W$ ($Z$) CR in the third (fourth) columns.
These prefit yields are the inputs for the statistical fit described below.
The samples are very pure, as the relative contribution of the $W$ ($Z$) CR is $95\%$ ($99\%$) from $W$ ($Z$) decays.
The definitions of the main normalizations parameters in the fit are
\vspace{-10pt}

{\small
  \begin{equation*}
    \begin{array}{lll}
      \big(B^\textsc{sr}_\smallW\big)_\text{estimate}
      &=\, N^\textsc{cr}_\smallW\cdot B^\textsc{sr}_\smallW/B^\textsc{cr}_\smallW
      &=\, B^\textsc{sr}_\smallW\cdot N^\textsc{cr}_\smallW/B^\textsc{cr}_\smallW
      \vspace{2pt}
      \\
      \big(B^\textsc{sr}_\smallZ\big)_\text{estimate}
      &=\, N^\textsc{cr}_\smallZ\cdot\underbrace{B^\textsc{sr}_\smallZ/B^\textsc{cr}_\smallZ}_{\alpha~\mbox{\scriptsize transfer}}
      &=\, B^\textsc{sr}_\smallZ\cdot\Bsp\!\!\underbrace{N^\textsc{cr}_\smallZ/B^\textsc{cr}_\smallZ}_{\beta~\mbox{\scriptsize normalization}},
    \end{array}
  \end{equation*}
}
\vspace{-8pt}

\noindent
where the event yields are for the observed data ($N$) and the MC estimate of the background ($B$).
The transfer factor $\alpha$ is the SR-to-CR ratio of the MC yields,
and is a quantity useful for visualizing how the systematic uncertainties partially cancel out.
The normalization $\beta$ is the data-to-MC ratio in the CR, which is extracted from the fit.
The analysis is performed in three $m_{jj}$ bins $i$, so $i$ also indexes $\alpha$ and $\beta$.

For the $W_{e\nu}^\textsc{high}$ subsample in the $W$ CR,
a yield parameter $\nu_\text{fake}$ is introduced to quantify the ``$e$ fakes,''
the group of electron candidates that are not prompt electrons.
This contamination occurs most often when a jet from a multijet event identified as an electron candidate.
The underlying idea is that the $W$ decays (multijets) have high (low) $\ETmiss$ resolution event-by-event.
Since $S_\textsc{met}$ is a proxy for $\ETmiss$ resolution,
a passing (failing) selection on $S_\textsc{met}{\,>\,}4\sqrt{\GeV}$
provides a $W_{e\nu}^\textsc{high}$ ($W_{e\nu}^\textsc{low}$) subsample depleted (enriched) in $e$ fakes.
In the fake-enriched $W_{e\nu}^\textsc{low}$ subsample, about a third of the events are due to $e$ fakes.
(For the $W_{e\nu}$ process, the $\ETmiss$ comes from the neutrino.
For this reason, the kinematic bias in $\ETmiss$ due to the $S_\textsc{met}$ selection was found to be negligible at the $1\%$ level.)
The resulting subsamples are tied together by a fixed ratio $\rho_\text{fake}$,
which is determined using a separate ``pure-fake'' region.

The pure-fake region ($F_{e\nu}$) is defined by a selection on the electron likelihood ($\mathcal{L}_e$).
Since $\mathcal{L}_e$ is optimized to separate electrons from backgrounds originating from dijet processes \cite{Aaboud:2016vfy},
requiring that the candidate's $\mathcal{L}_e$ value fail the \textsc{tight} definition \cite{ATLAS-CONF-2016-024},
while satisfying a looser definition, selects the $F_{e\nu}$ data sample.
As done above, the $S_\textsc{met}$ selection creates two subsamples $\big(F_{e\nu}^\textsc{high}$, $F_{e\nu}^\textsc{low}\big)$.
The $F_{e\nu}^\textsc{low}$-to-$F_{e\nu}^\textsc{high}$ ratio of the number of events in data is $\rho_\text{fake}$,
with the small amount of prompt $W$ contamination subtracted using MC.

Model testing uses a profile likelihood-ratio test statistic \cite{2011EPJC...71.1554C} in the CL$_\textsc{s}$-modified frequentist formalism \cite{CLs}.
The statistical treatment considers a total of $27$ bins: three $m_{jj}$ bins for each of nine subsamples
(one for the SR, four for the $W$ CR, two for the fake-enriched subsamples, two for the $Z$ CR).
A maximum-likelihood fit to the observed data in each $m_{jj}$ bin sets an upper limit,\footnote{%
  The likelihood is a product of Poisson functions,
  one for each sample of $N$ events while expecting $\lambda$,
  a Gaussian function for each systematic uncertainty, and
  a Poisson function for the number of MC events.
  In the simple scenario with only W and Z backgrounds,
  the $\lambda$ for the SR would be $S{\,+\,}\beta_\smallW{\,\cdot\,}B^\smallW_{\textsc{sr}}{\,+\,}\beta_\smallZ{\,\cdot\,}B^\smallZ_{\textsc{sr}}$,
  with each quantity multiplied by the response function for a systematic uncertainty.
  For the $W$ CR it is                      $\beta_\smallW{\,\cdot\,}B^\smallW_{\textsc{cr}}$ 
  and for the $Z$ CR it is                  $\beta_\smallZ{\,\cdot\,}B^\smallZ_{\textsc{cr}}$.
  See, e.g., Ref.\ \cite{ATLAS:2014aga}.
}
using a one-sided confidence level,
on $\mathcal{B}_\text{inv}$ for the $125\GeV$ Higgs boson and on the product $\sigma^\textsc{vbf}_\text{scalar}{\,\cdot\,}\mathcal{B}_\text{inv}$ for a scalar of different mass.
The prefit comparisons of data and MC are shown for all subsamples in Fig.~\ref{fig_01}.

The fit procedure extracts the nine floating parameters introduced above ($\beta_\smallW$, $\beta_\smallZ$, $\nu_\text{fake}$ for each $m_{jj}$ bin).
After the fit, the postfit $\beta$ parameters are consistent with the SM prefit prediction within their $1\,\sigma$ uncertainties.
The postfit comparisons of data and expected backgrounds are shown in Fig.~\ref{fig_02} for the two key variables, $m_{jj}$ and $\ETmiss$, for the $W$ and $Z$ CR.
The $m_{jj}$ ($\ETmiss$) plot groups the backgrounds to show the dependence of the distribution shape on the production mechanism (final state).

The postfit value of $\nu_\text{fake}$ (the product $\rho_\text{fake}{\,\cdot\,}\nu_\text{fake}$) is the absolute number of $e$ fake events in the $W_{e\nu}^\textsc{high}$ ($W_{e\nu}^\textsc{low}$) subsamples.
Since there is a $\nu_\text{fake}$ parameter for each bin $i$, the $m_{jj}$ shape is also predicted.
Apart from determining the $\rho_\text{fake}$ value, which is fixed in the fit, $F_{e\nu}$ is not part of the fit model.
We note that the $W_{e\nu}^\textsc{high}$-$W_{e\nu}^\textsc{low}$ samples are split by charge,
because $W^\pm$ production is not symmetric in $pp$ collisions.
However, the same $\nu_\text{fake}$ parameter is used for both charges because the $e$ fakes are expected to be symmetric in charge since they originate mostly from multijet events.

\begin{figure}[bt!]
  \begin{center}
  \includegraphics[width=1.0\columnwidth]{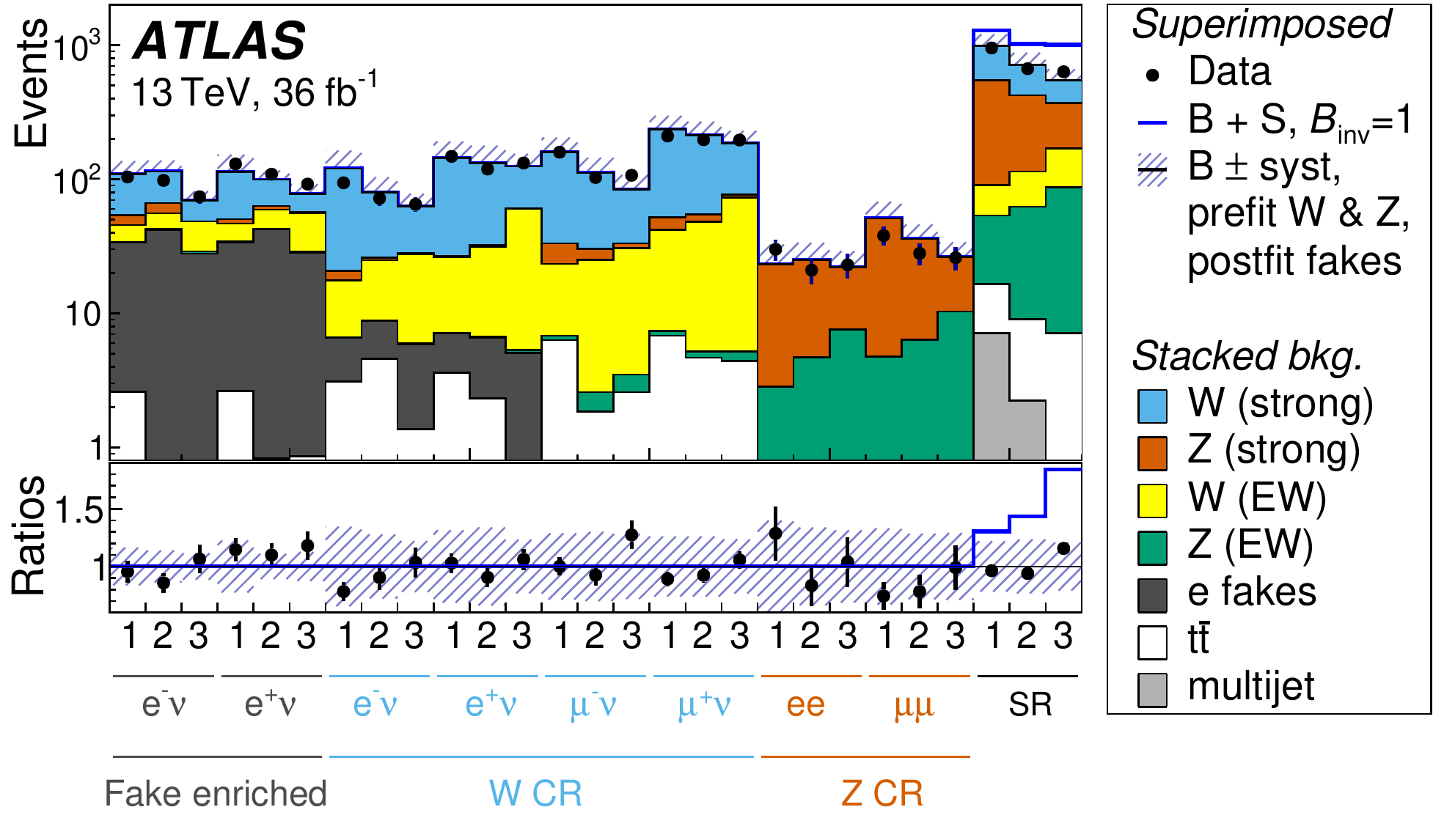}
  \end{center}
  \vspace{-6pt}
  \caption{
    Data-to-MC yield comparisons in the $27$ subsamples used in the statistical fit.
    The observed data $N$ (dots) are superimposed on the \textit{prefit} backgrounds $B$
    (stacked histogram with shaded systematic uncertainty bands).
    The hypothetical signal $S$ (empty blue histogram) is shown on top of $B$ for $\mathcal{B}_\text{inv}{\,=\,}1$.
    The bottom panels show the ratios of $N$ (dots) and $B{\,+\,S}$ (blue line) to $B$ with the systematic uncertainty band shown on the line at $1$.
    The $1$, $2$, and $3$ bin labels corresponds to 
    $1{\,<\,}m_{jj}{\,\le\,}1.5\TeV$,
    $1.5{\,<\,}m_{jj}{\,\le\,}2\TeV$, and
    $m_{jj}{\,>\,}2\TeV$, respectively.
    The ``$e$ fakes'' refers to $S_\textsc{met}{\,<\,}4\sqrt{\GeV}$ selection and is determined by the fit, so \textit{postfit} values are shown for the purposes of illustration.
    The diboson contribution is included in the electroweak (EW) $W$ and $Z$ bosons.
    \label{fig_01}
  }
\end{figure}

\begin{figure}[bt!]
  \begin{center}
  \includegraphics[width=0.8\columnwidth]{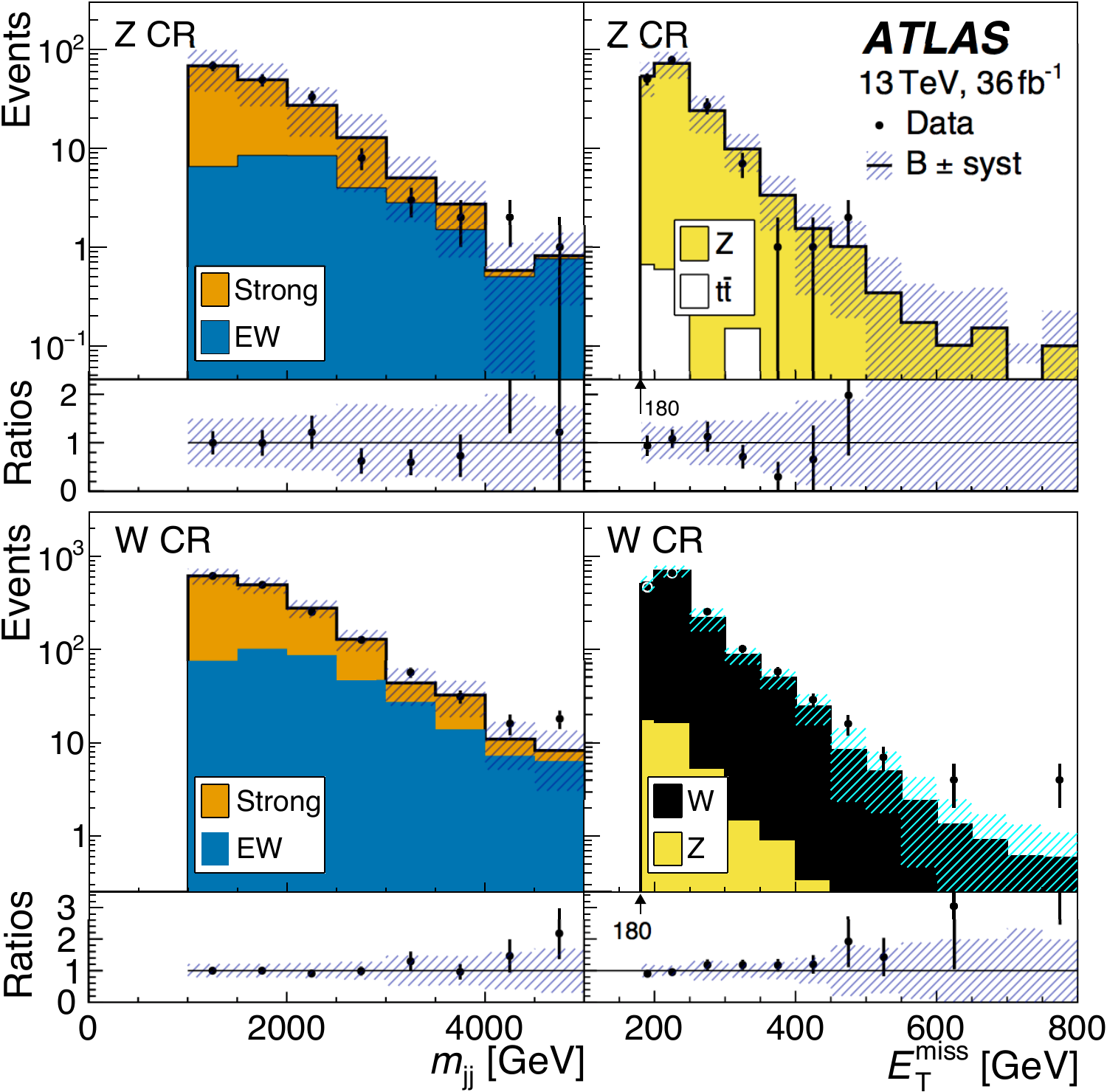}\\
  \end{center}
  \caption{
    Distribution of event yields in the $Z$ (top) and $W$ (bottom) control regions.
    The \textit{postfit} normalizations for $m_{jj}$ (left) and $\ETmiss$ (right) are summed over the subsamples.
    The $\ETmiss$ distributions start at $180\GeV$ as indicated.
    The observed data $N$ (dots) are superimposed on the sum of the backgrounds $B$ (stacked histogram with shaded systematic uncertainty bands).
    The breakdown of the $B$ is given in the lower left box in each panel.
    The bottom panels show the ratios of $N$ to $B$ with the systematic uncertainty band shown on the line at $1$.
    The ``other,'' as listed in Table~\ref{tab_01}, contribute a few events at low values of $m_{jj}$ and $\ETmiss$, and are omitted.
    The last bin in each plot contains the overflow.
    \label{fig_02}
  }
\end{figure}

The remaining processes---top quarks, dibosons, multijets---contribute negligibly to the SR (called ``other'' in Table~\ref{tab_01}).
The first two are estimated with MC using nominal cross sections.
The multijet contribution is very small, but it is a difficult process to estimate.
It is a potentially dangerous background because those events that pass the $\ETmiss$ selection are mostly due to instrumental effects.

The billionfold-or-more reduction of multijets after the event selection makes it impractical to simulate,
so a data-driven method based on a rebalance-and-smear strategy \cite{Aad:2012fqa} is used.
The assumption is that the $\ETmiss$ is due to jet mismeasurement in the detector response to jets and neutrinos from heavy-flavor decays \cite{Chatrchyan:2014lfa,CMS:2011ida}.
Using the jet-triggered sample,
the jet momenta are rebalanced by a kinematic fit, within their experimental uncertainties,
to obtain the balanced value of the jets' $\pT$.
The rebalanced jets are smeared according to jet response templates,
which are obtained from MC and validated with dijet data.
The rebalance-and-smear method predicts both the shape of the $\ETmiss$ distribution and the absolute normalization.
The procedure is verified in a $\Delta\phi_{jj}$-sideband validation region (VR) with $95\%$ purity of QCD multijet events.
This VR is defined by $1.8{\,<\,}|\,\Delta\phi_{jj}\,|{\,<\,}2.7$ 
and the loosening of the other requirements
($|\,\Delta\eta_{jj}\,|{\,>\,}3$,
$m_{jj}{\,>\,}0.6\TeV$, and
allow a third leading jet with $25{\,<\,}\pT{\,<\,}50\GeV$,
but no other jets with $\pT{\,>\,}25\GeV$).
The comparison of the predictions and the data in the VR shows good agreement (Fig.~\ref{fig_03}).
The multijet component is obtained using the rebalance-and-smear method with the associated systematic uncertainty bands,
while the non-multijet components are obtained using MC.

\begin{figure}[bt!]
  \begin{center}
  \includegraphics[width=0.8\columnwidth]{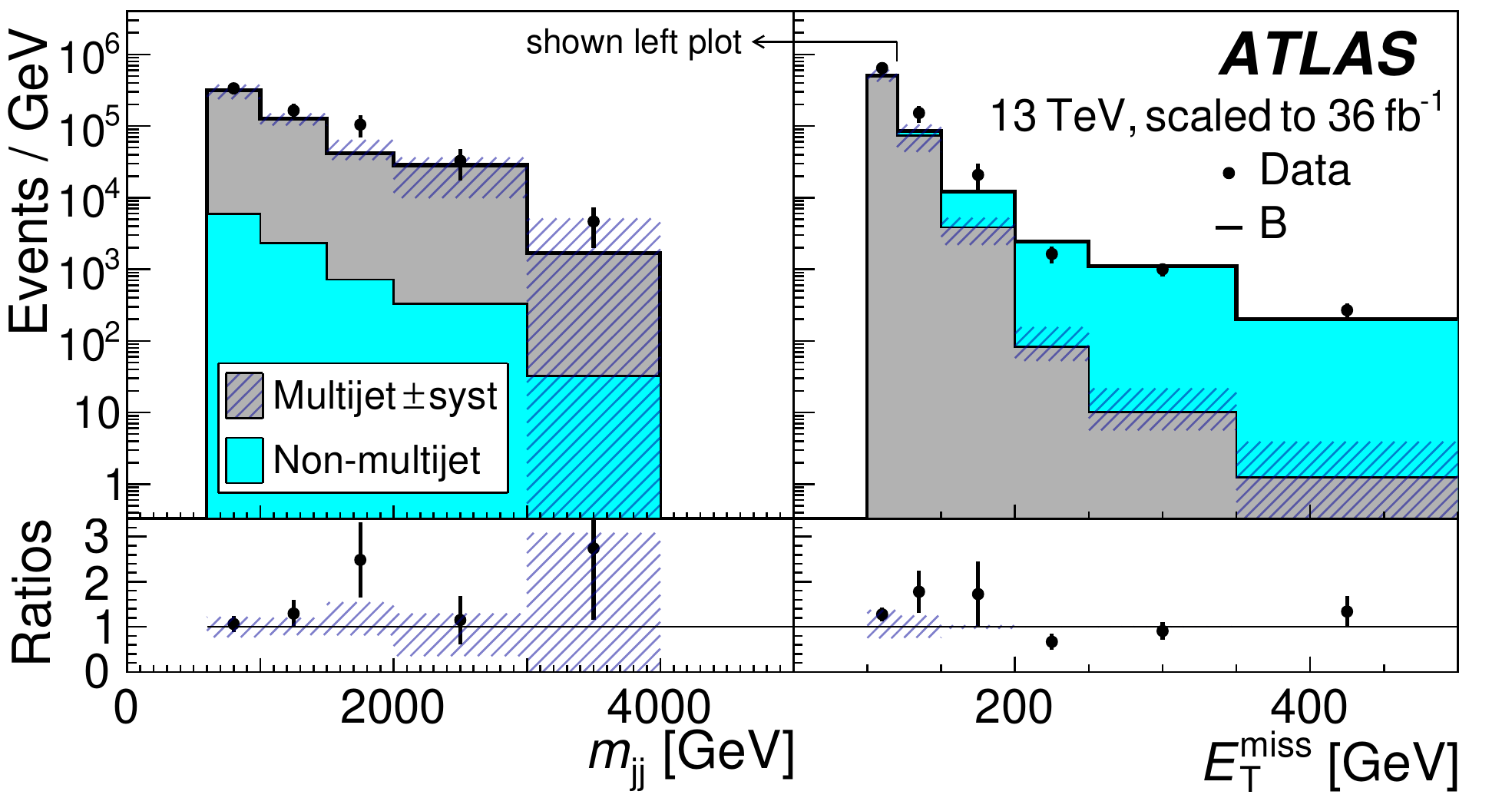}
  \end{center}
  \caption{
    Distribution of event yields in the multijet validation region for $m_{jj}$ (left) and $\ETmiss$ (right).
    The $m_{jj}$ plot shows the $100{\,<\,}\ETmiss{\,<\,}120\GeV$ subset of the right plot as indicated by the arrow.
    The $N$ observed data (dots) are superimposed on the sum of the $B$ backgrounds (stacked histogram).
    The systematic uncertainty band applies only to the multijet component.
    The statistical uncertainties are relatively large because of the weighting of the trigger samples with large prescale values.
    See the caption of Fig.~\ref{fig_02} for other plotting details.
    \label{fig_03}
  }
\end{figure}

\section{Uncertainties}\label{sec:syst}

Experimental and theoretical sources of uncertainties as well as the correlations between the various sources are described.
The resulting impact of the uncertainties on the yields and on the signal sensitivity is summarized later in Table~\ref{tab_02}.

Experimental sources of uncertainty are due mainly to the jet energy scale and resolution \cite{Aaboud:2017jcu},
$\ETmiss$ soft term \cite{Aaboud:2018tkc},
and lepton measurements \cite{ATLAS-CONF-2016-024,Aad:2016jkr}.
In order to reduce fluctuations due to limited MC sample size,
the uncertainties in number of expected events for the variations of jet energy scale and resolution for the strong and electroweak background samples are averaged.
This is motivated by the similarities of the kinematics and the detector effects for the two production processes for each $m_{jj}$ bin.
The uncertainty related to lepton identification or veto has a non-negligible (negligible) effect on $\alpha_\smallW$ ($\alpha_\smallZ$) because of the following scenarios.
The $W_{\ell\nu}$ background is significant in the SR, which results in an uncertainty for the cases related to the lepton veto.
The $Z_{\ell\ell}$ background is negligible in the SR, because the selection requires there to be no leptons.

The following experimental sources have small or negligible impact in the final result.
The pileup distribution and luminosity \cite{Aaboud:2016hhf,LUCID2} have a relatively small impact.
The trigger efficiency modeling, for both the lepton triggers for the CR and $\ETmiss$ triggers for the SR, are not listed in Table~\ref{tab_02}.
Their impact on the events yields was at the $1\%$ level and their impact on the signal sensitivity are found to be negligible.

Theoretical sources of uncertainty are due mainly to scale choices in fixed-order matrix-element calculations.
For the background processes,
QCD scales are varied for the resummation scale (resum.), renormalization scale (renorm.), factorization scale (fact.), and \textsc{ckkw} matching scale.
The first three scales in the list---technically called $q^2$, $\muR$, $\muF$, respectively---are varied by a factor of two \cite{Bothmann:2016nao,Heinemeyer:2013tqa}.
For the \textsc{ckkw} matching scale between the matrix element and the parton shower \cite{Gleisberg:2008ta},
the central value and the considered variations are $20^{+10}_{-5}\GeV$.
The higher-order electroweak corrections to the strongly produced $W$ or $Z$ are found to be negligible.

The effects of the theoretical variations are evaluated with a sample of generated MC events prior to reconstruction, which is larger than the reconstructed sample.
Moreover, in order to reduce fluctuations due to limited MC statistics,
the effect of the resummation and \textsc{ckkw} variations as a function of $m_{jj}$ are determined by a linear fit, using $m_{jj}$ values below the selection for the SR and a sample with loosened selection on $\Delta\eta_{jj}$ and $\Delta\phi_{jj}$.
We verified that an additional systematic uncertainty associated with the extrapolation is dominated by the statistical fluctuations in the varied samples.

For both signal and background, the effects of the choice of a parton distribution function (PDF) set have a relatively small impact.
The variations are considered using an ensemble of PDFs within the \textsc{nnpdf} set~\cite{Ball:2014uwa} and
the standard deviation of the distribution is taken as the uncertainty.

For the signal process,
the effect of the scale uncertainty on the third-jet veto for the gluon fusion plus two-jet contribution is evaluated using the jet-bin method \cite{Stewart:2011cf}.
The similar effect for the VBF contribution is evaluated by comparing the scale varied samples before and after the third-jet veto.
The impact on the Higgs signal yield is dominated by the VBF contribution, which is around $7\%$.

Statistical uncertainties are due to the data and MC sample sizes.

Systematic uncertainties are assumed to be either fully correlated or uncorrelated.
The uncertainties from the following sources in each independent $m_{jj}$ bin are correlated between the SR and CR: QCD scales, PDF, and lepton measurements.
The theoretical uncertainties due to QCD scales are uncorrelated between the following pairs:
signal vs.\ background, electroweak vs.\ strong production, and $W$ vs.\ $Z$ production.
Theoretical uncertainties are fully uncorrelated between bins of $m_{jj}$,
while the experimental uncertainties are fully correlated,
both of which are expected to be conservative assumptions. 

One major difference between Ref.~\cite{Aad:2015txa} and this paper---%
with the former (latter) employing (not employing) the $W$-to-$Z$ extrapolation strategy---%
is that we now have a larger $Z_{\ell\ell}$ control sample.
We found that the final limit result based on the statistical uncertainty of the enlarged $Z_{\ell\ell}$ control sample is similar to the result assuming the theoretical uncertainties on the $W$-to-$Z$ ratio
(including the associated MC sample statistical uncertainties).
This being the case, this paper adopts the method that is less dependent on theoretical assumptions.

\begin{table}[tb!]
{\small
\caption{
  Sources of uncertainty.
  The first set shows $\Delta$, the \emph{relative} improvement of the $95\%$ CL upper limit on $\mathcal{B}_\text{inv}$ when the source of uncertainty is ``removed'' by fixing it to its best-fit value.
  The ``visual'' column shows bars whose lengths from the center tick are proportional to $\Delta$.
  The second set shows the effect on the yields and the $\alpha$ transfer factors for the $1{\,<\,}m_{jj}{\,\le\,}1.5\TeV$ bin.
  The yields are for the signal process in the SR ($S$), $Z$ MC in the SR $\big(B_\smallZ^\textsc{sr}\big)$, and $Z$ MC in the CR $\big(B_\smallZ^\textsc{cr}\big)$.
  The $\alpha_\smallZ$ is given to demonstrate the reduction in the uncertainty in the ratio $B_\smallZ^\textsc{sr}/B_\smallZ^\textsc{cr}$.
  The individual yields for the $W$ are not shown because the cancellation effects are similar to the $Z$ counterparts.
  The value for ``$3^\text{rd}$ jet veto'' corresponds only to the uncertainty related to jet bin migration for signal processes;
  the corresponding effect for the background processes are evaluated in the various jet energy and theoretical variations.
  The abbreviations for the theoretical sources are described in the text.
  The `-' indicates that the quantity is not applicable.
  The ``combined'' rows at the bottom are not simple sums of the rows above because of the $\Delta$ metric;
  the symbols $(\dag, \ddag, \star)$ are parenthetically defined in the table.
  The penultimate (last) row shows the summary impact of removing the systematic uncertainties due to the experimental and theoretical sources (as well as statistical uncertainties of the MC samples).
  \label{tab_02}
}
\begin{center}
\begin{tabular*}{0.55\columnwidth}{l ccp{0pt} cccccp{0pt}}
  \dbline
  \multirow{1}{*}{Source}  
  & \multicolumn{3}{l}{\Bsp $\mathcal{B}_\text{inv}$\,improve.\,[\%]\BSP} & \multicolumn{6}{l}{\Bsp Yields,\,$\alpha$\,changes\,(\%)}\\
  & \multicolumn{3}{l}{\Bsp using\,all\,$m_{jj}$\,bins                  } & \multicolumn{6}{l}{\Bsp in $1{<}m_{jj}{\le}1.5\TeV$} \vspace{2pt}\\
  \cline{2-3}\cline{5-9}
  \phantom{$X^{\mbox{Y}}$}
  &\!\asp{$\Delta$}
  &   \BSP visual \BSP
  &
  &   \asp{$S$}
  &   \asp{$B_\textsc{sr}^\smallZ$}
  &   \asp{$B_\textsc{cr}^\smallZ$}
  &   \asp{$\alpha_\smallZ$}
  &   \asp{$\alpha_\smallW$}\bsp
  \\
  \sgline
  \balkenscale{18}{-40.0}
  Experimental ($\dagger$)
  \\ ~~Jet energy scale                      \bsp&\!\asp{10} &\Bsp{\color{myblue}\Balkenx{10}{10}{10}{0}{0}} &&\asp{12} &\asp{ 7} &\asp{ 8} &\asp{ 8} &\asp{6}
  \\ ~~Jet energy resol.                     \bsp&\!\asp{ 2} &\Bsp{\color{myblue}\Balkenx{10}{ 2}{ 2}{0}{0}} &&\asp{ 2} &\asp{ 0} &\asp{ 1} &\asp{ 1} &\asp{4}
  \\ ~~$\ETmiss$\,soft\,term                 \bsp&\!\asp{ 1} &\Bsp{\color{myblue}\Balkenx{10}{ 1}{ 1}{0}{0}} &&\asp{ 2} &\asp{ 2} &\asp{ 2} &\asp{ 2} &\asp{2}
  \\ ~~Lepton id., veto                      \bsp&\!\asp{ 2} &\Bsp{\color{myblue}\Balkenx{10}{ 2}{ 2}{0}{0}} &&\asp{ -} &\asp{ -} &\asp{ -} &\asp{ 0} &\asp{4}
  \\ ~~Pileup distrib.                       \bsp&\!\asp{ 1} &\Bsp{\color{myblue}\Balkenx{10}{ 1}{ 1}{0}{0}} &&\asp{ 3} &\asp{ 1} &\asp{ 2} &\asp{ 3} &\asp{1}
  \\ ~~Luminosity                            \bsp&\!\asp{ 0} &\Bsp{\color{myblue}\Balkenx{10}{ 0}{ 0}{0}{0}} &&\asp{ 2} &\asp{ 2} &\asp{ 2} &\asp{ -} &\asp{-}
  \\
  \sgline
  Theoretical ($\ddagger$)
  \\ ~~Resum.\ scale\bsp                     \bsp&\!\asp{ 1} &\Bsp{\color{myblue}\Balkenx{10}{ 1}{ 1}{0}{0}} &&\asp{ -} &\asp{ 2} &\asp{ 3} &\asp{ 0} &\asp{2}
  \\ ~~Renorm., fact.                        \bsp&\!\asp{ 2} &\Bsp{\color{myblue}\Balkenx{10}{ 2}{ 2}{0}{0}} &&\asp{ -} &\asp{20} &\asp{19} &\asp{ 1} &\asp{2}
  \\ ~~\textsc{ckkw} matching                \bsp&\!\asp{ 4} &\Bsp{\color{myblue}\Balkenx{10}{ 4}{ 4}{0}{0}} &&\asp{ -} &\asp{ 2} &\asp{ 3} &\asp{ 1} &\asp{5}
  \\ ~~PDF                                   \bsp&\!\asp{ 0} &\Bsp{\color{myblue}\Balkenx{10}{ 0}{ 0}{0}{0}} &&\asp{ 1} &\asp{ 1} &\asp{ 2} &\asp{ 1} &\asp{1}
  \\ ~~3$^\text{rd}$ jet veto                \bsp&\!\asp{ 2} &\Bsp{\color{myblue}\Balkenx{10}{ 2}{ 2}{0}{0}} &&\asp{ 7} &\asp{ -} &\asp{ -} &\asp{ -} &\asp{-}
  \\
  \sgline
  Statistical
  \\ ~~MC sample ($\star$)                   \bsp&\!\asp{12} &\Bsp{\color{myblue}\Balkenx{10}{12}{12}{0}{0}} &&\asp{ 4} &\asp{ 5} &\asp{ 9} &\asp{10} &\asp{9}
  \\ ~~Data sample                           \bsp&\!\asp{21} &\Bsp{\color{myblue}\Balkenx{10}{27}{27}{0}{0}} &&\asp{ 6} &\asp{ 5} &\asp{12} &\asp{12} &\asp{6}
  \\
  \sgline
  \multicolumn{3}{l}{Combined}
  \\ ~~All $\dagger$ sources                 \bsp&\!\asp{17} &\Bsp{\color{mygold}\Balkenx{10}{17}{17}{0}{0}}
  \\ ~~All $\ddagger$ sources                \bsp&\!\asp{10} &\Bsp{\color{mygold}\Balkenx{10}{10}{10}{0}{0}}
  \\ ~~Combine $\dagger$, $\ddagger$         \bsp&\!\asp{28} &\Bsp{\color{mygold}\Balkenx{10}{28}{28}{0}{0}}
  \\ ~~Combine $\dagger$, $\ddagger$, $\star$\bsp&\!\asp{42} &\Bsp{\color{mygold}\Balkenx{10}{42}{42}{0}{0}}
  \\ \dbline
\end{tabular*}
\end{center}
}
\end{table}

The sources of uncertainty are grouped into the three main categories given above (Table~\ref{tab_02}).
The impact of each source is measured in two ways:
(1) on the $95\%$ CL upper limit on $\mathcal{B}_\text{inv}$ and
(2) on the event yields and $\alpha$ transfer factors.
Impact (1) assesses the percentage improvement of the $\mathcal{B}_\text{inv}$ limit if that source of uncertainty is removed
after fixing the associated parameter to its best-fit value.
Impact (2) demonstrates that the systematic uncertainties in the individual yields partially cancel out for many of the theoretical sources.
However, for many of the experimental sources the cancellation is not achieved due to limited MC statistics of the varied samples.
For example, the effects of varying the renormalization and factorization scales change the MC yield in the $Z$ SR $\big(B_\textsc{sr}^\smallZ$ in Table~\ref{tab_02}$\big)$ and the $Z$ CR $\big(B_\textsc{cr}^\smallZ\big)$
by about {$20\%$}, but the $\alpha_\smallZ$ transfer factor changes by only {$1\%$.}
In Table~\ref{tab_02}, only the $1{\,<\,}m_{jj}{\,\le\,}1.5\TeV$ yields are shown for the purpose of illustrating the partial cancellation in the ratio.

In general, the uncertainties are higher with $m_{jj}$.
The MC sample statistics is the largest source of systematic uncertainties,
with the uncertainty increasing with $m_{jj}$ due to limited number of simulated events.
The theory uncertainties are also higher with $m_{jj}$ values for the same reason.
The experimental jet energy uncertainties are also affected by the limited sample size,
with larger fluctuations because of fluctuations that do not cancel for each individual systematic variations.
For the sources contributing the largest uncertainties,
the $\alpha_\smallZ$ and the $\alpha_\smallW$ variations in the three $m_{jj}$ bins are shown graphically in Fig.~\ref{fig_04}.

\begin{figure}[bt!]
  \begin{center}
  \includegraphics[width=0.8\columnwidth]{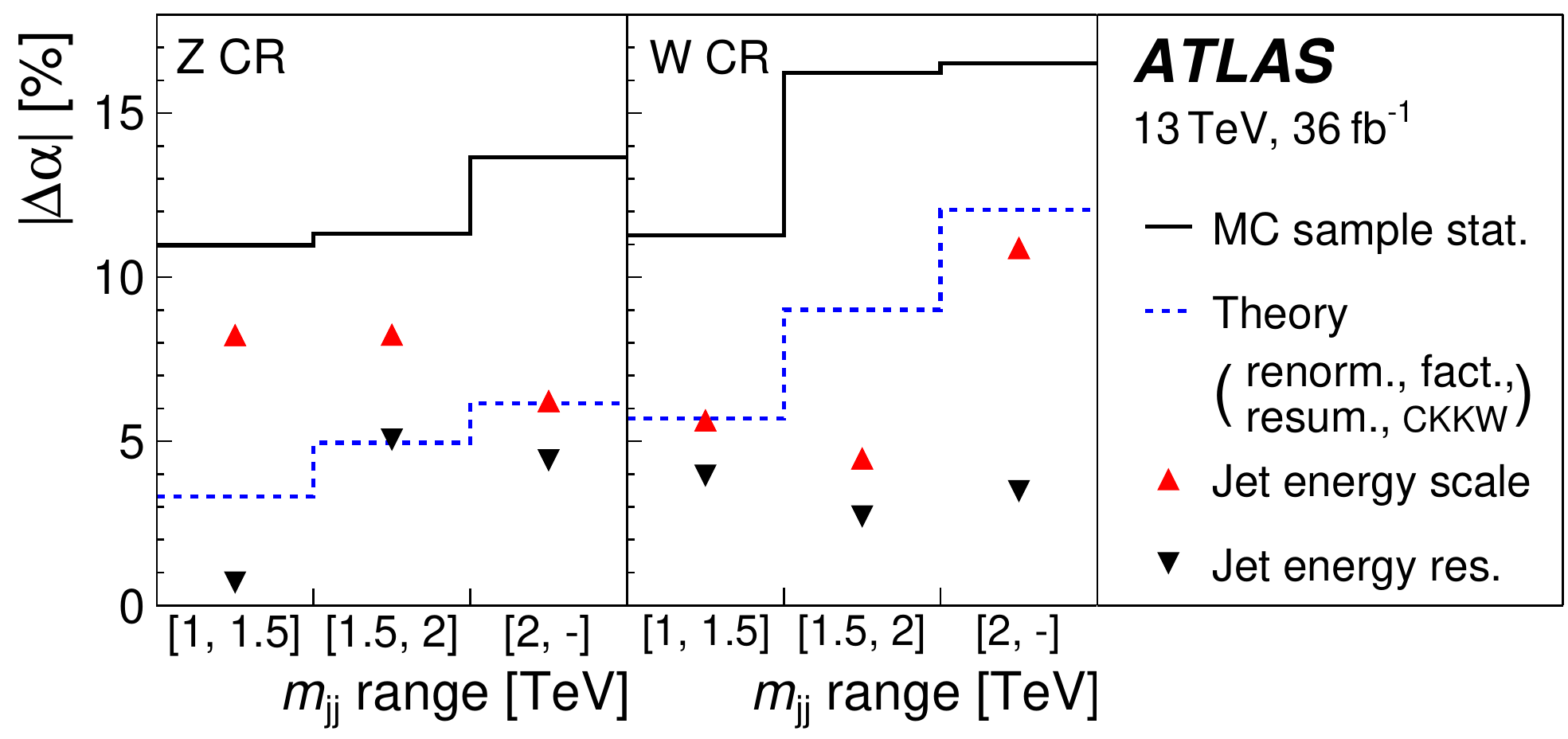}
  \end{center}
  \vspace{-6pt}
  \caption{
    Contributions to the relative uncertainty in the transfer factors $\alpha_\smallZ$ (left) and $\alpha_\smallW$ (right) in the three $m_{jj}$ bins of the SR.
    The theoretical uncertainties from the sources noted in the legend are combined in quadrature.
    \label{fig_04}
  }
\end{figure}

The combination of uncertainties from various sources shows that the dominant category has a systematic origin (penultimate row of Table~\ref{tab_02}).
The lack of MC statistical precision for background processes with $m_{jj}{\,>\,}2\TeV$ has the largest impact on $\mathcal{B}_\text{inv}$.
We note that the $\Delta$ values are percent improvements of the final limit on $\mathcal{B}_\text{inv}$,
so they do not add in quadrature or in any such standard statistical combinations.

\begin{figure}[bt!]
  \begin{center}
  \includegraphics[width=0.8\columnwidth]{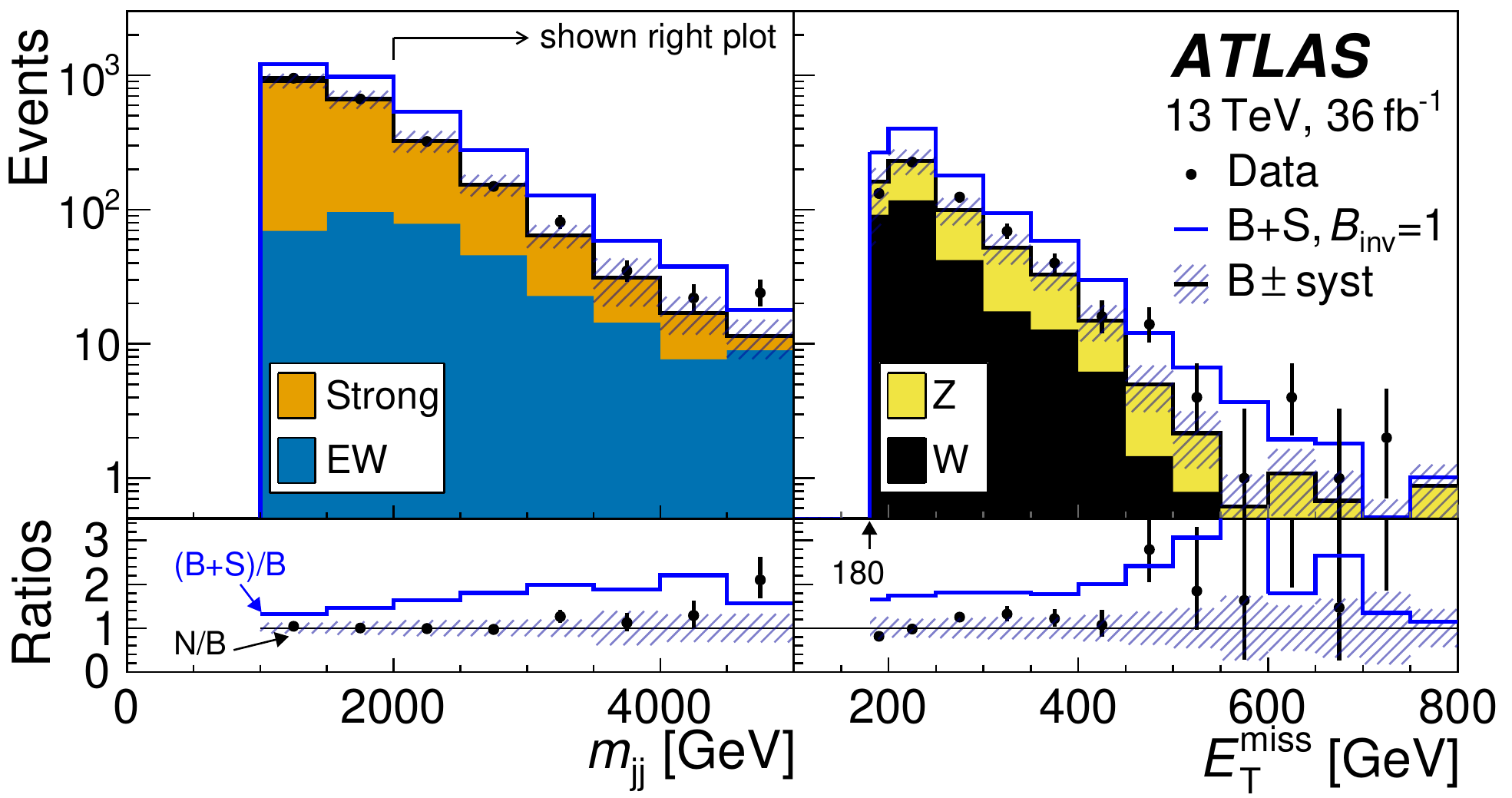}
  \end{center}
  \vspace{-6pt}
  \caption{
    Distribution of event yields in the signal region for $m_{jj}$ (left) and $\ETmiss$ (right).
    The $\ETmiss$ distributions start at $180\GeV$ and shows the most sensitive $m_{jj}{\,>\,}2\TeV$ subset of the SR as indicated by the arrow.
    The \textit{postfit} normalizations for $m_{jj}$ ($\ETmiss$)
    distributions use separate background, $B$, normalizations in the 
    three (one) $m_{jj}$ bins of $1{\,<\,}m_{jj}{\,\le\,}1.5\TeV$, $1.5{\,<\,}m_{jj}{\,\le\,}2\TeV$, and $m_{jj}{\,>\,}2\TeV$ ($m_{jj}{\,>\,}2\TeV$).
    and sum the contributions from $W$ and $Z$ bosons (electroweak and strong production modes).
    The hypothetical signal $S$ (empty blue histogram) is shown on top of $B$ for $\mathcal{B}_\text{inv}{\,=\,}1$.
    The bottom panels show the ratios of $N$ (dots) and $B{\,+\,S}$ (blue line) to $B$ with the systematic uncertainty band shown on the line at $1$.
    The bin width in the $m_{jj}$ plots ($\ETmiss$) is $500\GeV$ ($50\GeV$ except for the first bin with the non-zero entry, which is $20\GeV$).
    See the caption of Fig.~\ref{fig_02} for other plotting details.
    \label{fig_05}
  }
\end{figure}

\section{Results and interpretations}\label{sec:results}

The $2252$ observed events in the SR are divided among the three $m_{jj}$ bins defined previously: $952$, $667$, and $633$ events.
These values are consistent with the background-only postfit yields of the sum of the background processes of $2100$ events,
which are divided among the three $m_{jj}$ bins:
$850{\,\pm\,}113$,
$660{\,\pm\,}90$, and
$590{\,\pm\,}81$, respectively.
The uncertainty represents the combined effect due to experimental and theoretical systematic uncertainties.
These postfit values are also consistent with the prefit predictions.
The expected signal yields (for $\mathcal{B}_\text{inv}{\,=\,}1$ for VBF and gluon fusion) are $300$, $310$, and $460$, respectively,
and the last $m_{jj}$ bin has the highest sensitivity with $S/B{\,\approx\,}0.8$.

The postfit SR event distributions of $m_{jj}$ and $\ETmiss$ are shown in Fig.~\ref{fig_05},
and we observe agreement, within uncertainties, between the data and the expected backgrounds.

Figure~\ref{fig_05}(a) also shows that the $S/B$ ratio rises with increasing $m_{jj}$ values, 
which motivates our division of the SR into multiple bins.
The total electroweak contribution in the SR is relatively small at $\mathcal{O}(10\%)$ (Table~\ref{tab_01}),
but the much flatter distribution of $m_{jj}$ makes it an important contribution to the final result.
As noted in Section~\ref{sec:cr}, the background estimation is done independently for each $m_{jj}$ bin to reduce the dependence on $m_{jj}$ modeling.

The fit, assuming the $125\GeV$ Higgs boson, gives the observed (expected) upper limit on $\mathcal{B}_\text{inv}$ of {$0.37$ $\big(0.28{\,}^{+0.11}_{-0.08}\big)$} at $95\%$ CL,
and {$0.32$ $\big(0.23{\,}^{+0.11}_{-0.10}\big)$} at $90\%$ CL,
where the uncertainties placed on the expected limit represent the $1\sigma$ variations.
With this result, connections to \textsc{wimp} dark matter can be made in the context of Higgs portal models \cite{higgs_portal}.
The limit on $\mathcal{B}_\text{inv}$ can be used to set limit on the Higgs-\textsc{wimp} coupling by the \textsc{wimp}-nucleon scattering cross section formulae ($\sigma_{\textsc{wimp}\text{-nucleon}}$).
In this paper, scalar and Majorana fermion \textsc{wimp} models are considered \cite{Eboli:2000ze,Fox:2011pm,deSimone:2014pda}.

The overlay of the interpretation of this result with the limits from some of the direct detection experiments \cite{Akerib:2016vxi,Cui:2017nnn,Aprile:2018dbl}
shows the complementarity in coverage (Fig.~\ref{fig_06}(a)).
For the scalar \textsc{wimp} interpretation cross sections are excluded at values ranging from $\mathcal{O}\big(10^{-42}\big)$ to $\mathcal{O}\big(10^{-45}\big)\,\text{cm}^2$ and
for the Majorana fermion \textsc{wimp} interpretation the exclusion range is from $\mathcal{O}\big(10^{-45}\big)$ to $\mathcal{O}\big(10^{-46}\big)\,\text{cm}^2$,
depending on the \textsc{wimp} mass.
The uncertainty band in the plot uses an updated computation of the nucleon form factors \cite{Hoferichter:2017olk}.

The correlation between $\mathcal{B}_\text{inv}$ and $\sigma_{\textsc{wimp}\text{-nucleon}}$ is presented in the effective field theory framework
assuming that the new-physics scale is $\mathcal{O}(1)\TeV$ \cite{Aad:2015txa}, well above the scale probed at the Higgs boson mass.
Adding a renormalizable mechanism for generating the fermion \textsc{wimp} masses could modify the above-mentioned correlation \cite{Baek:2014jga}.

\begin{figure}[bt!]
  \begin{center}
  \begin{overpic}[width=0.6\columnwidth]{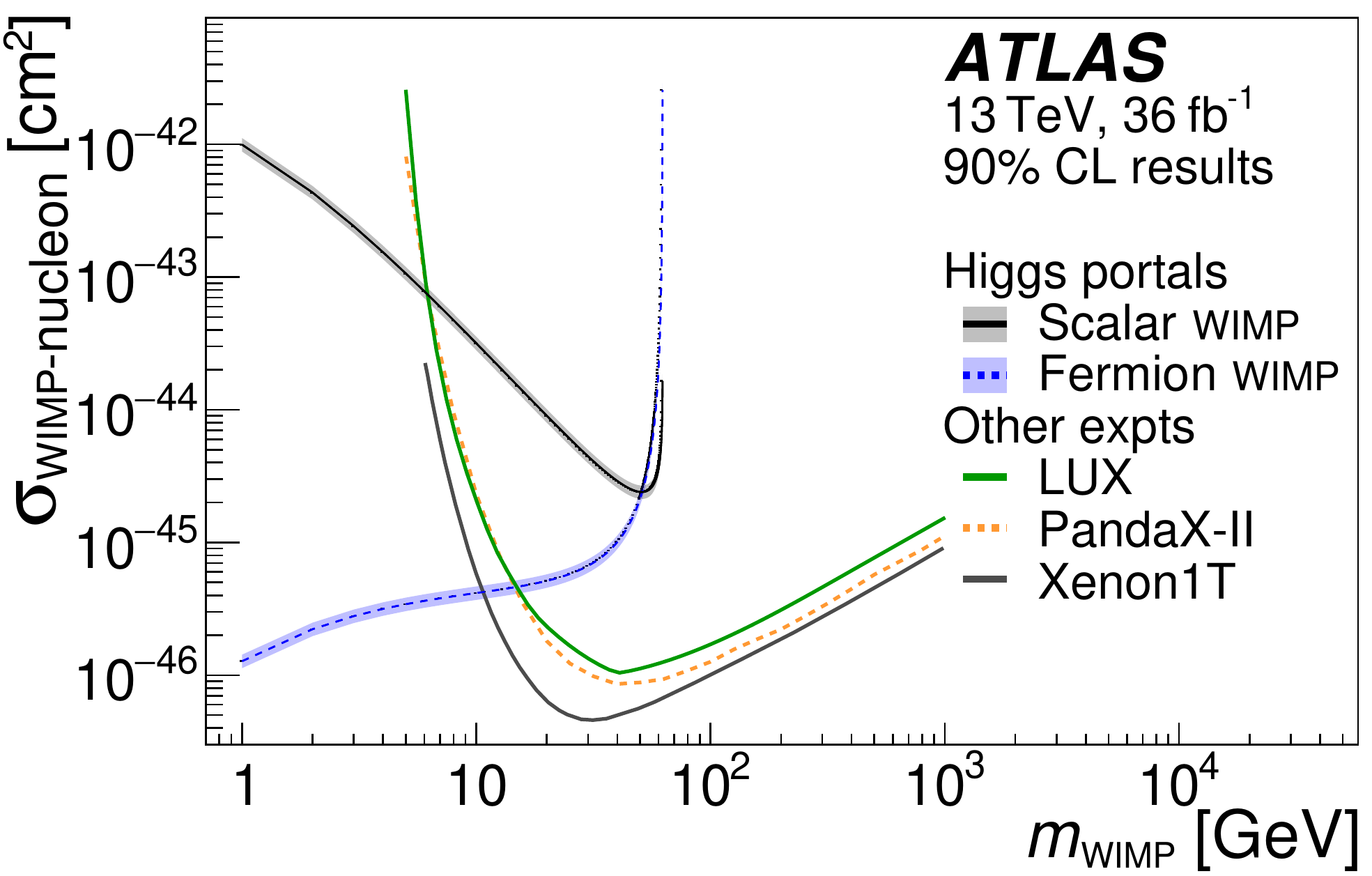}   \put(-12,35){\footnotesize (a)}\end{overpic} \\
  \begin{overpic}[width=0.6\columnwidth]{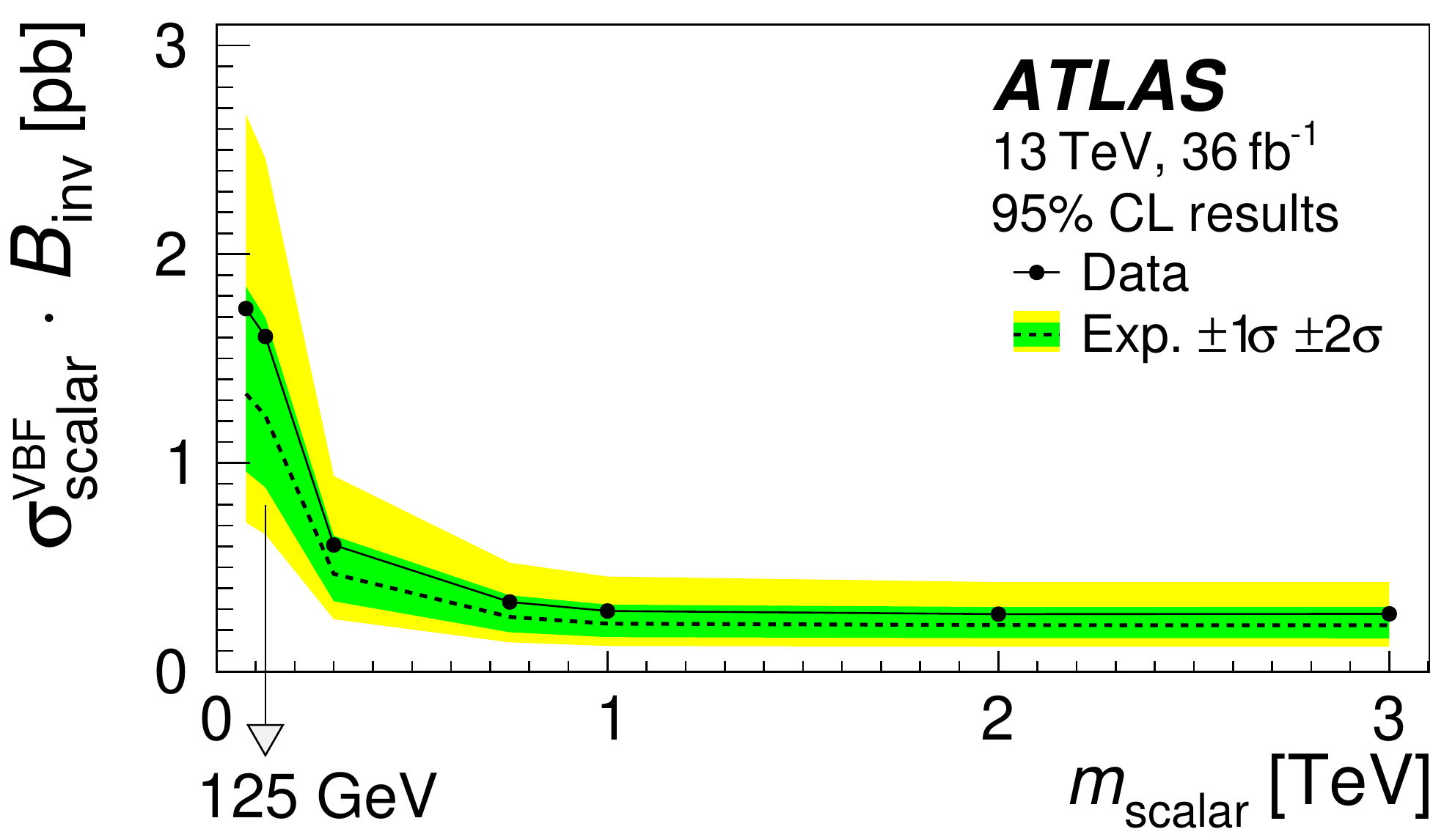} \put(-12,35){\footnotesize (b)}\end{overpic}
  \end{center}
  \vspace{-12pt}%
  \caption{
    Upper limits on (a) the spin-independent \textsc{wimp}--nucleon cross section using Higgs portal interpretations of $\mathcal{B}_\text{inv}$ at $90\%$ CL 
    vs.\ $m_\textsc{wimp}$ and
    (b) the VBF cross section times the branching fraction to invisible decays at $95\%$ CL vs.\ $m_\text{scalar}$.
    The top plot shows results from Ref.~\cite{Akerib:2016vxi,Cui:2017nnn,Aprile:2018dbl}.
  }
  \label{fig_06}
\end{figure}

In place of the $125\GeV$ Higgs boson, the same selection is applied to additional scalars with masses ($m_\text{scalar}$) of up to $3\TeV$ assuming only VBF production.
The fraction of VBF signal events that pass the signal region event selections corresponding to the acceptance times efficiency ranges from $0.6$--$3\%$.
The signal efficiency for the inclusive $m_{jj}{\,>\,}1\TeV$ selection increases with the mass of the scalar boson,
because the VBF jets is more forward with higher mass, and thus have more events at higher values of $m_{jj}$.
The limit on $\sigma^\textsc{vbf}_\text{scalar}{\,\cdot\,}\mathcal{B}_\text{inv}$ as a function of $m_\text{scalar}$ is shown in Fig.~\ref{fig_06}(b).
The $95\%$ confidence level upper limits on the cross section times branching fraction are in the range of $0.3$--$1.7\pb$.

\section{Conclusions}

A search for Higgs boson decays into invisible particles is presented using the $36.1\ifb$ of $pp$ collision data taken at $\sqrt{s}{\,=\,}13\TeV$
collected in 2015 and 2016 by the ATLAS detector at the LHC.
The targeted signature is the VBF topology with two energetic jets with a wide gap in $\eta$ and large $\ETmiss$.

Assuming the Standard Model cross section for the $125\GeV$ Higgs boson,
an upper limit of $0.37$ is set on $\mathcal{B}_\text{inv}$ at $95\%$ CL.
This result is interpreted using Higgs portal models to exclude regions in the $\sigma_{\textsc{wimp}\text{-nucleon}}$ vs.\ $m_\textsc{wimp}$ parameter space to 
exclude cross section values ranging from $\mathcal{O}\big(10^{-42}\big)$ to $\mathcal{O}\big(10^{-46}\big)\,\text{cm}^2$,
depending on the \textsc{wimp} mass and the \textsc{wimp} model.

Searches for invisible decays of scalars with masses of up to $3\TeV$ are reported for the first time from ATLAS in the VBF production mode.
These results are rather general and can be used for further interpretations.

\FloatBarrier

\clearpage
\section*{Acknowledgements}

% Acknowledgements for papers with collision data
% Version 24-Oct-2018

% Standard acknowledgements start here
%----------------------------------------------
We thank CERN for the very successful operation of the LHC, as well as the
support staff from our institutions without whom ATLAS could not be
operated efficiently.

We acknowledge the support of ANPCyT, Argentina; YerPhI, Armenia; ARC, Australia; BMWFW and FWF, Austria; ANAS, Azerbaijan; SSTC, Belarus; CNPq and FAPESP, Brazil; NSERC, NRC and CFI, Canada; CERN; CONICYT, Chile; CAS, MOST and NSFC, China; COLCIENCIAS, Colombia; MSMT CR, MPO CR and VSC CR, Czech Republic; DNRF and DNSRC, Denmark; IN2P3-CNRS, CEA-DRF/IRFU, France; SRNSFG, Georgia; BMBF, HGF, and MPG, Germany; GSRT, Greece; RGC, Hong Kong SAR, China; ISF and Benoziyo Center, Israel; INFN, Italy; MEXT and JSPS, Japan; CNRST, Morocco; NWO, Netherlands; RCN, Norway; MNiSW and NCN, Poland; FCT, Portugal; MNE/IFA, Romania; MES of Russia and NRC KI, Russian Federation; JINR; MESTD, Serbia; MSSR, Slovakia; ARRS and MIZ\v{S}, Slovenia; DST/NRF, South Africa; MINECO, Spain; SRC and Wallenberg Foundation, Sweden; SERI, SNSF and Cantons of Bern and Geneva, Switzerland; MOST, Taiwan; TAEK, Turkey; STFC, United Kingdom; DOE and NSF, United States of America. In addition, individual groups and members have received support from BCKDF, CANARIE, CRC and Compute Canada, Canada; COST, ERC, ERDF, Horizon 2020, and Marie Sk{\l}odowska-Curie Actions, European Union; Investissements d' Avenir Labex and Idex, ANR, France; DFG and AvH Foundation, Germany; Herakleitos, Thales and Aristeia programmes co-financed by EU-ESF and the Greek NSRF, Greece; BSF-NSF and GIF, Israel; CERCA Programme Generalitat de Catalunya, Spain; The Royal Society and Leverhulme Trust, United Kingdom. 

The crucial computing support from all WLCG partners is acknowledged gratefully, in particular from CERN, the ATLAS Tier-1 facilities at TRIUMF (Canada), NDGF (Denmark, Norway, Sweden), CC-IN2P3 (France), KIT/GridKA (Germany), INFN-CNAF (Italy), NL-T1 (Netherlands), PIC (Spain), ASGC (Taiwan), RAL (UK) and BNL (USA), the Tier-2 facilities worldwide and large non-WLCG resource providers. Major contributors of computing resources are listed in Ref.~\cite{ATL-GEN-PUB-2016-002}.
%----------------------------------------------

\clearpage
\printbibliography

\clearpage % ATLAS Collaboration author list
% Reference date of EXOT-2016-37 is 2018-03-08
% Author list last updated on date 10-MAY-19
% Data extracted on 10-May-2019 for paper reference EXOT-2016-37
% at 4:51pm
 
\begin{flushleft}
{\Large The ATLAS Collaboration}

\bigskip

M.~Aaboud$^\textrm{\scriptsize 35d}$,    
G.~Aad$^\textrm{\scriptsize 100}$,    
B.~Abbott$^\textrm{\scriptsize 127}$,    
O.~Abdinov$^\textrm{\scriptsize 13,*}$,    
B.~Abeloos$^\textrm{\scriptsize 131}$,    
D.K.~Abhayasinghe$^\textrm{\scriptsize 92}$,    
S.H.~Abidi$^\textrm{\scriptsize 166}$,    
O.S.~AbouZeid$^\textrm{\scriptsize 40}$,    
N.L.~Abraham$^\textrm{\scriptsize 155}$,    
H.~Abramowicz$^\textrm{\scriptsize 160}$
% ...
\bigskip
\\

% ...
$^{13}$Institute of Physics, Azerbaijan Academy of Sciences, Baku; Azerbaijan.\\
% ...
$^{35}$$^{(a)}$Facult\'e des Sciences Ain Chock, R\'eseau Universitaire de Physique des Hautes Energies - Universit\'e Hassan II, Casablanca;$^{(b)}$Facult\'{e} des Sciences, Universit\'{e} Ibn-Tofail, K\'{e}nitra;$^{(c)}$Facult\'e des Sciences Semlalia, Universit\'e Cadi Ayyad, LPHEA-Marrakech;$^{(d)}$Facult\'e des Sciences, Universit\'e Mohamed Premier and LPTPM, Oujda;$^{(e)}$Facult\'e des sciences, Universit\'e Mohammed V, Rabat; Morocco.\\
% ...
$^{40}$Niels Bohr Institute, University of Copenhagen, Copenhagen; Denmark.\\
% ...
$^{92}$Department of Physics, Royal Holloway University of London, Egham; United Kingdom.\\
% ...
$^{100}$CPPM, Aix-Marseille Universit\'e, CNRS/IN2P3, Marseille; France.\\
% ...
$^{127}$Homer L. Dodge Department of Physics and Astronomy, University of Oklahoma, Norman OK; United States of America.\\
% ...
$^{131}$LAL, Universit\'e Paris-Sud, CNRS/IN2P3, Universit\'e Paris-Saclay, Orsay; France.\\
% ...
$^{155}$Department of Physics and Astronomy, University of Sussex, Brighton; United Kingdom.\\
% ...
$^{160}$Raymond and Beverly Sackler School of Physics and Astronomy, Tel Aviv University, Tel Aviv; Israel.\\
% ...
$^{166}$Department of Physics, University of Toronto, Toronto ON; Canada.\\
% ...
$^{a}$ Also at Borough of Manhattan Community College, City University of New York, New York NY; United States of America.\\
$^{b}$ Also at Centre for High Performance Computing, CSIR Campus, Rosebank, Cape Town; South Africa.\\
$^{c}$ Also at CERN, Geneva; Switzerland.\\
$^{d}$ Also at CPPM, Aix-Marseille Universit\'e, CNRS/IN2P3, Marseille; France.\\
$^{e}$ Also at D\'epartement de Physique Nucl\'eaire et Corpusculaire, Universit\'e de Gen\`eve, Gen\`eve; Switzerland.\\
% ...
$^{*}$ Deceased

\end{flushleft}

% Created with Glance <Atlas.Glance@cern.ch>

\end{document}